\documentclass[sigconf]{acmart}
\usepackage{CJKutf8}

\AtBeginDocument{%
  }

\copyrightyear{2026}
\acmYear{2026}
\makeatletter
\def\@ACM@copyright@check@cc{}
\makeatother
\setcopyright{cc}
\setcctype{by-nc-nd}
\acmConference[CHI '26]{Proceedings of the 2026 CHI Conference on Human Factors in Computing Systems}{April 13--17, 2026}{Barcelona, Spain}
\acmBooktitle{Proceedings of the 2026 CHI Conference on Human Factors in Computing Systems (CHI '26), April 13--17, 2026, Barcelona, Spain}
\acmPrice{}
\acmDOI{10.1145/3772318.3790583}
\acmISBN{979-8-4007-2278-3/2026/04}

\begin{document}

\title[]{The State's Politics of "Fake Data"}

\author{Chuncheng Liu}
\affiliation{
\department{Department of Communication Studies}
\institution{Northeastern University}
\country{Boston, MA, USA}
}
\email{ch.liu@northeastern.edu}

\author{danah boyd}
\affiliation{
\department{Department of Communication}
\institution{Cornell University}
\country{Ithaca, NY, USA}
}
\email{dmb478@cornell.edu}

% \renewcommand{\shortauthors}{Liu and boyd}

\begin{abstract}
Data have power. As such, most discussions of data  presume that records should mirror some idealized ground truth. Deviations are viewed as failure. Drawing on two ethnographic studies of state data-making\textemdash in a Chinese street-level bureaucrat agency and at the US Census Bureau\textemdash we show how seemingly “fake” state data perform institutional work. We map four moments in which actors negotiate between representational accuracy and organizational imperatives: creation, correction, collusion, and augmentation. Bureaucrats routinely privilege what data do over what they represent, creating fictions that serve civil servants' self-interest and enable constrained administrations. We argue that “fakeness” of state data is relational (context dependent), processual (emerging through workflows), and performative (brought into being through labeling and practice). We urge practitioners to center fitness-for-purpose in assessments of data and contextual governance. Rather than chasing impossible representational accuracy, sociotechnical systems should render the politics of useful fictions visible, contestable, and accountable.
\end{abstract}




\maketitle

\section{Introduction}
\begin{figure}[h]
    \centering
    \includegraphics[width=1\linewidth]{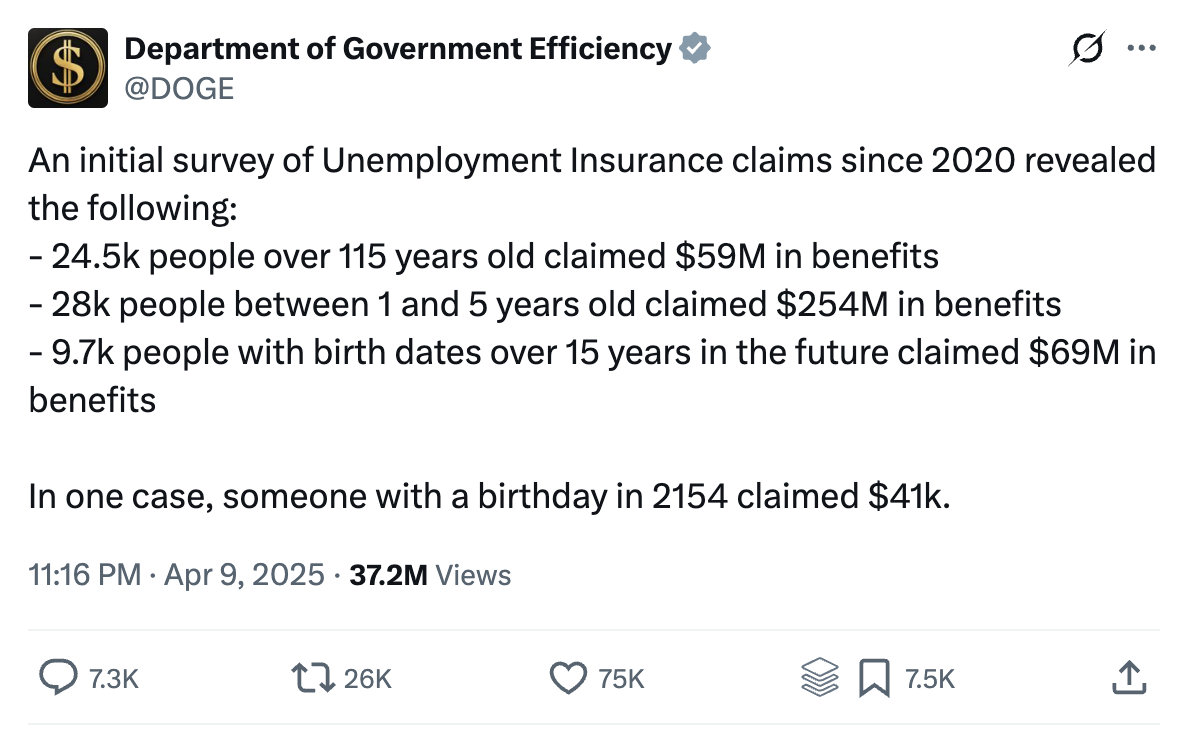}
    \caption{A screenshot of an April 9, 2025 social media post from the Department of Government Efficiency (DOGE) }
    \label{fig:tweet}
\end{figure}

On April 9, 2025, the US Department of Government Efficiency (DOGE) posted on X (the social media platform) that tens of thousands of people were claiming government benefits despite being over 115 years old, under the age of five, or born 15+ years in the future (Figure \ref{fig:tweet}). Fifteen minutes later, Elon Musk\textemdash who was leading DOGE at the time\textemdash retweeted DOGE's post, noting “Your tax dollars were going to pay fraudulent unemployment claims for fake people born in the future!”  

After civil servants started to respond, a different narrative emerged. The so-called “fake” people reflected in these records were not receiving government benefits; rather, these data entries were intentionally created for administrative purposes. They were “pseudo claim records” to support anti-fraud efforts encouraged by the US Department of Labor where “the implausibility of the records was the point.” Facing mounting allegations, a state agency spokesperson pushed back, arguing that DOGE was “misunderstanding this data at best, mischaracterizing this data at worst” \cite{badger_musks_2025}.

This exchange reveals more than mere political theater. It exemplifies how debates over state data typically unfold through a representationalist framework that proves inadequate for understanding how state data function. This framework relies on a binary understanding where “real” data accurately represent external reality and are inherently good, while “fake” data fail at representation and signal corruption or incompetence \cite{crooks_representationalism_2017, kim_epistemologies_2024}. DOGE’s social media attack reflected the dominance of this perspective, presuming that the public would expect birth dates in government databases to correspond to actual births, ages to reflect biological facts, and  records labeled people to always reflect identifiable individuals. 

However\textemdash as STS, HCI, and critical data studies scholars have increasingly shown\textemdash seemingly anomalous data can fulfill critical administrative functions, from maintaining system compatibility across technological transitions to processing exceptional cases that fall outside standard categories \cite{lampland_false_2010,bowker_sorting_2000,cheong_theorizing_2023,Hoeyer_2023}. The designation of such data as “fake” is a political act that obscures the complex institutional work that these data perform. To investigate these dynamics, three questions motivate our research: 

\begin{itemize}
\item RQ1: How do actors at different positions in state data pipelines define, label, and mobilize “fake data,” and what does this labeling accomplish?
\item RQ2: Through what organizational moments does a data record’s status shift between “real” and “fake,” and what work does each shift perform for the system?
\item RQ3: Under what conditions do bureaucrats prioritize what data do over what they represent, and how are “acceptable fictions” negotiated, bounded, and maintained in practice?
\end{itemize}

To answer these questions, we draw on ethnographic research from two distinct sociopolitical contexts. The first case explores how street-level bureaucrats in Chinese government agencies document neighborhood volunteering activities. The second investigates the US Census Bureau’s data production processes, which directly affect political representation and resource allocation. Together, we identify four critical moments – creation, correction, collusion, and augmentation – through which the fakeness of data is produced, negotiated, and mobilized toward varying ends. These illustrative cases and moments reveal the institutional and organizational logics for normalizing “fake data” production. They also show how the boundaries between “real” and “fake” data are continuously negotiated by multiple actors with divergent stakes and interpretive frameworks. 

Our theoretical intervention reframes “fake data” as a relational, processual, and performative category rather than an intrinsic property of records. “Fake” is often deployed as a political maneuver to undermine the legitimacy of data and the agencies that produce them. Rather than trying to tease out which categories of data warrant the label “fake,” we seek to reclaim the label and reveal the logics underlying the wide array of practices that prompt actors to claim fakeness.

By moving beyond the “fake/real” data binary and proposing the concept of “fakeness” of data, we advance three central arguments. First, fakeness of data is relational: whether data are taken as real or fake depends on actors' positionality in specific contexts and power relationships. Second, fakeness of data is processual: it must be understood within the state’s organizational structure of data production pipeline. Third, fakeness of data performs specific functions: those involved in data production often accept seemingly “fake” data because they prioritize what data \textit{do} over what they \textit{represent}. Taken together, our reframing engages with the posthumanist performativity theory to redirect attention to the work fake data perform – for whom, when, and with what consequences. 

Bridging STS, HCI, and critical data studies to examine the politics of state data, this paper offers distinct contributions. By rejecting the representational focus that has long centered debates about state data, we bring organizational structures and institutional workflows to the fore through new conceptual vocabularies. Empirically, we contribute to HCI’s relatively limited engagement with public-sector contexts (compared to civil society and industry) by providing a cross-national, inside-the-bureaucracy ethnographic approach that traces the pipeline by which the fakeness of data are normalized and put to work. Finally, we translate these insights into practical suggestions for designing systems and policies that move toward governance frameworks capable of managing necessary uncertainties rather than pursuing impossible standards of perfect representation. 

In the following sections, we first review the literature on the politics of state data and fake data. After detailing our ethnographic methods and cases, we analyze the production of data fakeness through four interconnected moments. We conclude by arguing that recognizing the inevitability and functionality of “fake” data is essential for understanding and improving how states govern through data to avoid the empty promise of data-driven governance that only create new vulnerabilities for state legitimacy.

\section{Related Work}

\subsection{The State’s Fake Data Problem}
Data and the state have always been entangled. States have long been the primary producers of large-scale data crucial to governance, creating data sets such as the census and gross domestic product (GDP). Statistics originated as “the science of the state,” and what philosopher Ian Hacking called the “avalanche of printed numbers” in the 19th century was directly associated with the expansion of state power \cite{hacking1990taming}. Additionally, the state relies on data to function; as anthropologist James Scott famously argued, the state’s capacity to impose social order rests on its ability to “see” through data collection \cite{scott_seeing_1999}. In recent years, scholars have argued that the high-modernist states Scott described have evolved into dataist states in the era of big data \cite{liu_seeing_2022,liu_social_2022,fourcade_learning_2020, farrell2023moral}. Such states view individuals as “data doubles” \cite{Bouk_2018} and society as an ocean of data flows, claiming the responsibility to collect, analyze, and model these flows to govern. 

The public is implicated in state data at multiple levels. At the population level, data have become essential for the state to organize and distribute social, economic, and political resources to different communities \cite{anderson_american_2015,bouk_democracys_2022}. At the individual level, one’s position in a state database can affect one's citizenship, access to social welfare, likelihood of criminalization, or right to mobility during pandemic lockdown \cite{Brayne_Christin_2021,chen_maintainers_2023,cheong_theorizing_2023,lageson_digital_2020,moon_datafication_2025, singh_seeing_2021}.

Given this centrality, much work on the history of state data examines how governments perform objectivity and neutrality to render data “real” and trustworthy \cite{micheli_emerging_2020,porter_trust_1996,johns_governance_2021}. Consequently, “fake data” typically appear in literature as a problem to be solved. Scandals and controversies are read as signs of accountability failures or corruption, prompting efforts to identify, correct, and prevent such records in the name of better data-driven governance \cite{hansen_what_2012,lageson_digital_2020,vetro_data_2021}. Political scientists, for instance, have documented how the Chinese state adopted a quantified vision that unintentionally created a system where numbers define politics, resulting in a limited state scope characterized by data fabrication in various domains \cite{ghosh_making_2020,wallace_seeking_2022}. Similarly, during the COVID-19 pandemic, scholars documented the systematic manipulation of death statistics across regime types where suspiciously low variations in death reporting failed standard statistical tests for natural number distribution \cite{annaka_political_2021}. Together, these studies exemplify a problem-oriented approach: developing detection techniques, advocating for transparency mechanisms, and calling for international verification to ensure the accuracy of state data.

HCI has engaged state data along a similar trajectory. Interest in how public-sector data are made and used has grown \cite{nelimarkka_what_2025,lee_reconfiguring_2024}. Much of this work examines facilitators and barriers to the digitization of state administration (e.g. \cite{holtgrave_qualitative_2025}), inclusive design for democratic engagement from civic-tech perspectives (e.g. \cite{kim_factful_2015}), and the affordances of government digital platforms (e.g. \cite{lu_learning_2020}). By contrast, comparatively fewer studies center government data; those that do often show how erroneous public-sector datasets – used to make eviction decisions \cite{tran_situating_2024} or to train predictive-policing algorithms \cite{roskin-frazee_bureaucratic_2025} – produce social harms and inequalities. In a design-improvement vein, HCI researchers have also developed systems to improve access to government data (e.g. \cite{kim_factful_2015}) and to detect and flag errors (e.g. \cite{tran_situating_2024}). 

The problem-oriented approach to fake data has generated crucial empirical findings across diverse political contexts, but it perpetuates a narrow understanding of data as either succeeding or failing at representation, “adopt[ing] an ontologically simplistic view of the politics and nature of statistics” \cite{aragao_many_2022}. The limitations of this representationalist framework become apparent when we examine how data actually function within state apparatuses. Not all data with representational inaccuracies are labeled as fake, and not all data pejoratively labeled as fake have representational inaccuracies. These theoretical and empirical tensions call for a new perspective on the state’s politics of fake data.

\subsection{Problems with the Fake Data Problem}
By putting the problem-oriented lens of fake data and the state in dialogue with critical data studies and humanistic HCI research, we can identify three related problems. 

The first problem is an ontological one, which understands the state’s fake data as “false data” purely in terms of its representational accuracy, often in a binary way. “Real” and “fake” are “trouser words.” They derive meaning not from their own properties but through contrast with their opposites \cite{austin_sense_1964}. Labeling data as “fake” presupposes an authentic alternative, and vice versa. Yet this binary obscures how data often lack an “external reality from which they are ontologically distinct” \cite{desrosieres_how_2001} that might never exist to be accurately represented. Most data function as “ways of assessing and managing uncertainty” \cite{eyal_crisis_2019}. State-produced data frequently rely on historically contingent and contested categories – such as GDP measures – that do not represent real objects outside their constructed worlds \cite{mitchell_rule_2002,espeland_sociology_2008}. Even when it comes to data that aim to represent “real” objects, such as the number of people in a space at a time or the amount of grain in a field, it is often impossible to perfectly produce what relevant actors would agree were “real” data due to limits in state capacity, technical challenges, and messy social dynamics \cite{desrosieres_how_2001,ghosh_making_2020,Garfinkel_Bittner_1967}. In a representational sense, almost no data are truly real.

Once we loosen the grip of a strict real/fake binary, a second problem\textemdash a processual one\textemdash emerges: fake data tend to be characterized as “fixed data,” with their production reduced to discrete moments of creation or manipulation. The primary focus of statistical manipulation has been on how street-level bureaucrats fabricate data at the ground level \cite{blundo_dealing_2006}, or on how technocrats edit economic data from the statistical yearbook database \cite{aragao_many_2022}. Yet this perspective often treats data work as a one-time translation from reality to representation, neglecting how it is an ongoing processes that continuously reshapes the composition of data within specific organizational structures \cite{feinberg_design_2017,winthereik_data_2024}. In other words, this perspective treats data as either “raw” or “cooked,” ignoring how data are co-constituted by stakeholders, devices, and infrastructures across myriad social contexts \cite{bowker2008memory}. As critical data studies scholars have shown, data undergo constant transformation through aggregation, disaggregation, reclassification, and reinterpretation at different administrative levels \cite{garnett_air_2017,gordon_data_2024,ammitzboll_flugge_street-level_2021}. Both the representational accuracy of data and their practical work evolve dynamically, serving different actors’ shifting purposes across institutional contexts while shifting both the meaning of data and their relationship to “real” or “fake.” 

The third problem is the normative assumption that “fake data” are merely “failed data”: deviations to be corrected once inaccuracies are detected. This view overlooks how data serve diverse purposes beyond representation, and fails to explain why practices deemed “fake” become routinized or even preferred within state institutions. Here, we draw on Garfinkel and Bittner's foundational insight into “good organizational reasons for bad clinical records” \cite{Garfinkel_Bittner_1967}, where they argued that seemingly deficient data records serve organizational needs. Lampland’s study of Hungarian collective farms, for example, shows that false numbers were valued by superiors as signs of compliance, where “ensuring the relative accuracy of the numerical sign becomes secondary to mastering the logic of formal procedures” \cite{lampland_false_2010}. Kim and colleagues similarly demonstrate, in their analysis of COVID-19 dashboards, how officials prioritized keeping systems operational amid uncertainty, navigating multiple epistemologies of data and addressing missing information beyond questions of representation \cite{kim_epistemologies_2024}. Studies of other organizations observe a similar pattern. For example, Pine and Mazmanian trace how institutional logics of “safety through systems” generate perfect-looking, yet factually inaccurate, hospital data \cite{pine_institutional_2014}, while Seberger and Gupta argue that divergences between data representations and lived experience should not be treated as deficits to repair, but as openings for dialogue \cite{seberger_designing_2025}. Across these cases, the value of fake data is situated: social positions, incentives, and infrastructural constraints can render “fake” data more functional than purportedly “real” data. 

\subsection{A Critical Lens on Fakeness of Data}
We conceptualize the politics of fake data from the standpoint of the STS theory of performativity. As Karen Barad argues, this theoretical lens “shifts the focus from questions of correspondence between descriptions and reality to matters of practices/doings/actions” \cite{barad2003posthumanist}. Our approach aligns with posthumanist frameworks that treat sociotechnical systems not as “fixed representations of entities, but only exist in their situated intra-action” \cite{Frauenberger2019}, entangling human and non-human actors, institutional infrastructures, and organizational demands.

From this standpoint, “fake data” are not records with an intrinsic property of falsity. Instead, fake data are process-oriented objects continually acted upon, reshaped, and mobilized within specific organizational contexts. This framework builds on scholarship that destabilizes the ontological status of data by placing them within broader sociotechnical and institutional arrangements \cite{paparova2023,winthereik_data_2024}. Applying this to our inquiry, “fake data” is therefore an umbrella term capturing a family resemblance of practices through which data diverge from their ostensible referents—sometimes intentionally, sometimes inadvertently, sometimes productively. The analytic questions shift from  \textit{what makes data fake} to \textit{what work does fakeness do} and \textit{how do actors accomplish fakeness in practice}.

Our focus on “fakeness” represents an intentional move away from binary categorizations. We treat data’s representational capacity as relational and processual, shifting across time, scale, and institutional position. This perspective illuminates how diverse forms of divergence—fabrication, augmentation, inference, misclassification, category mismatch, interface-driven error, and procedural performance—operate within state data pipelines and acquire meaning in relation to bureaucratic power and positionality. Within different organizational configurations, fakeness performs different functions, while its legitimacy is constructed, contested, negotiated, and mobilized. Such dynamics do not imply randomness; as our analysis shows, the reconfiguration of fakeness follows patterned logics anchored in bureaucratic hierarchy, accountability structures, and political expectations.

Our research identifies concrete mechanisms and articulates critical moments through which fakeness is enacted and performed. In doing so, we show how data’s divergence from reality is not simply a deviation to be pathologized, but a constitutive feature of state sociotechnical systems—one that reveals how power, practice, and infrastructure co-produce the conditions under which the state “sees” and acts.

\section{Methods}
As ethnographers, we had been independently studying government workers involved in Chinese state data and the 2020 US census prior to this collaboration. In discussing our fieldwork, we started to identify commonalities and differences between our cases. This prompted us to consider which aspects of our data could complement and complicate each other.

From our broader fieldwork efforts, we selected the Chinese volunteering data project and the US census’ data production process to provide a strategic comparative analysis. Beyond the evident contrast in political regimes, these two cases offer a productive divergence in institutional objectives. Our cases highlight different stages of the bureaucratic lifecycle: the Chinese case emphasizes the social negotiation of data at the street level, while the US case emphasizes the statistical processing at the aggregate level. Together, they reveal different facets of how fakeness serves state data production. 

Despite their various differences, government workers in both cases share a foundational principle: they prioritize  functional validity over representational accuracy. Our comparison demonstrates that the normalization of “fake data” is not unique to authoritarian or democratic bureaucracies but is a structural inevitability of state legibility projects where the work data must do supersedes the reality they are meant to reflect. In this way, our research also echoes recent calls to view China and the US as mirrors, revealing “patterns that can also be observed elsewhere—sometimes more sharply—so that parallels, not just contrasts, come into view” \cite{liu2025china}.

\subsection{Community Volunteering Data in China}
The Chinese government has invested significantly in intervening in social life and shaping citizens’ moral subjectivities. The New Era Civilization Practice (NECP) campaign, launched in 2018, is one of its latest developments \cite{trauth-goik_civilized_2023}. Among various activities, NECP encourages local governments to organize citizens to conduct voluntary activities, such as cleaning the street or making dumplings for the elderly in the neighborhood \cite{hu_building_2023}. Most local governments assigned the task of organizing these volunteering activities to grassroots committees, the semi-state agencies working in different neighborhoods. Grassroots committees are not formal state agencies but work closely with the residents to complete the state’s tasks such as data collection and service delivery \cite{tang_grid_2020,chen_maintainers_2023}. In urban areas, they are residents’ committees; in rural areas, they are called villagers’ committees. 

From 2019 to 2021, the first author spent ten months in “Meritown,” an anonymized northern Chinese city, conducting a project on quantitative governance at three grassroots committees in one urban neighborhood and two villages. Liu worked in each office for three months on a daily basis, shadowing local street-level bureaucrats to learn how they organize events, interact with residents, and upload event data. He also conducted 108 semi-structured interviews with bureaucrats, volunteers, and ordinary residents in the communities. The research project was reviewed and approved by the Institutional Review Board (IRB) of the University of California San Diego.

Meritown is one of the pioneers in promoting volunteering activities, receiving positive recognition from the national media and the central government. In Meritown, a valid data entry for a volunteering activity must meet three standards: 1) involve more than 15 participants; 2) last for more than 90 minutes, and 3) provide evidence to prove the activity’s occurrence, including a group picture of all the participants, two pictures showing volunteers working, and a digital check-in and check-out record from the volunteers’ smart city mobile phone application. The Meritown municipal government claimed that more than 3,000 events were organized every month. Among its population of 700,000 people, about 30\% of Meritown's residents have registered volunteering records, resulting in more than six million hours of volunteering activities.

\subsection{2020 Census in the United States}
In 2020, the United States conducted its 24th decennial census. The data produced through this effort are constitutionally configured to apportion political power and federal dollars \cite{anderson_american_2015}. This particular census was notably fraught, shaped by the COVID-19 pandemic, political interference, and natural disasters \cite{sullivan_assessing_2023}. Given the context in which the census took place and changes to its methodology, many outside the Census Bureau were especially wary of the quality of the data and the legitimacy of the bureau's actions \cite{boyd_differential_2022,Rocco_2025,Nanayakkara_Hullman_2023}. 

The second author conducted a four-year ethnographic study of the 2020 census \cite{boyd2026_data_made}. While embedded inside the Census Bureau, boyd observed government officials contending with the data that they received from both self-response mechanisms and the fieldwork conducted by temporary workers known as enumerators. She observed several groups, including multiple teams dedicated to “quality assurance” and “fraud detection.” These teams helped determine whether data collected by enumerators or through self-response were sufficiently “real” to commit to record. One of these teams was authorized to send questionable data out for re-collection, but they had to be judicious in doing so given the resources involved. Other teams had the ability to leverage different “editing” and “processing” techniques to repair data that seemed “wrong.” All sources of data collection create uncertainties and challenges for those tasked with rooting out problematic, erroneous, or peculiar data\cite{Groves_Lyberg_2010}. Yet, civil servants' job requires them to determine which data to keep, which to edit, and which to delete. 

This second author also participated in various external events where she observed how  census stakeholders\textemdash including civil rights advocates, community organizers, political operatives, and data users\textemdash interpreted the Census Bureau's work. From this vantage point, she was able to see how different stakeholders grappled with knowing whether the data were “good”\textemdash or, in some stakeholders’ language, “fair and accurate.” In multiple events, she heard stakeholders attack the bureau's work by declaring certain data to be fake. Notably, what was labeled fake by stakeholders varied significantly based on partisan affiliation and epistemic orientation. 

Complementing her ethnographic fieldwork, she also conducted 103 semi-structured interviews with government officials and external stakeholders in which she asked her subjects to speak to data quality. This study was reviewed and approved by Microsoft Research's IRB.

\subsection{Data Analysis}
All  personal and geographical information in this article (except for public figures) is anonymized to protect the study participants. For our analysis, we employed an abductive analytical approach that moved iteratively between empirical observations and theoretical development \cite{timmermans_theory_2012}. Our collaboration began when we began discussing how “manipulation” was a source of contestation in each of our field sites. After recognizing the similarities between our respective cases, both authors returned to their data and independently coded their respective field notes and interview transcripts. In the process, we began  noting moments when the status of data as “real” or “fake” was raised, contested, negotiated, or transformed. We initially turned to scholarship from STS and political sociology to anchor our analysis, which led us to consider how organizational and institutional contexts shape the routine production of “fake” data.

The two authors then met weekly for three months to share interview excerpts, fieldwork observations, and historical background from both cases, enabling systematic cross-case comparison. Despite the different political contexts and data types, we discovered striking parallels in how bureaucrats navigated tensions between representational accuracy and institutional functionality, even as their specific objectives diverged. Through these discussions, we developed processual maps that revealed four critical moments where the relationship of data to their purported referents were actively reconfigured: creation, correction, collusion, and augmentation. These moments subsequently provided the organizing structure for our analysis. 

This analytical process also allowed us to move beyond our initial engagement with statistical manipulation, which still implicitly treated data’s “realness” or “fakeness” as a question of correspondence with external reality. As we observed increasingly public and highly political debates in both countries about “fake data” that often targeted data that we would not ourselves classify as intentionally or accidentally fabricated, we began to reconceptualize fake data in our analysis. This shift enabled us to connect our empirical observations with critical data studies’ critiques of representationalism \cite{kim_epistemologies_2024,crooks_representationalism_2017,matzner_beyond_2016}, while articulating our distinct theoretical and empirical contributions to understanding how “fakeness” operates within state bureaucracies.

\section{Analysis}
Government workers are often aware that the data they produce are not representationally accurate. They are both open about—and accepting of—this. Across both field sites, we witnessed moments in which officials described their own outputs as “fake”\textemdash sometimes jokingly, sometimes with resignation or cynicism. On occasion, bureaucrats acknowledged intentional fabrication. More commonly, however, these practices were enacted as routine administrative work without recognition. When researchers pressed them on whether such practices might constitute “fake data” in the public’s eyes, bureaucrats frequently defended them as necessary, justified, or simply the only feasible way to make the system function.

Understanding such practices requires attending to the organizational processes through which data are transformed and rendered actionable within the state. The production of fake data unfolds through a series of interconnected moments as it moves through bureaucratic hierarchies. In what follows, we trace four such moments—creation, correction, collusion, and augmentation. Each involves distinct actors, constraints, and interpretive frames. Together, these moments reveal how bureaucrats negotiate between representational accuracy and institutional functionality, and how “fakeness” is produced, managed, validated, reshaped, and reinterpreted within state data systems.

\subsection{Creating Fakeness}
The initial moment of data creation is critical and typically falls to street-level bureaucrats who operate at the interface between the state apparatus and society. These frontline workers exercise considerable discretion in translating policy mandates into practice \cite{gordon_data_2024,ammitzboll_flugge_street-level_2021}. While data fabrication can involve outright falsification, the phenomenon is often more nuanced.

At the extreme end of the spectrum, state workers sometimes fabricate data entirely. The US census has historically struggled with the various ways in which “enumerators” (the temporary workers who knock on doors on behalf of the Census Bureau) create fictitious responses rather than conduct actual surveys \cite{bouk_democracys_2022}. One version of this practice is known as “curbstoning,” which describes moments when workers allegedly sat on curbs inventing household data. The Census Bureau punishes workers who are caught fabricating data; these data are then thrown out.

Most fake data, however, occupy a spectrum between authentic and invented elements. In other words, they contain different degrees of fakeness. To see the spectrum, it is important to recognize the role that motivation plays. Census enumerators sometimes inflate population counts for political or economic gain, adding people who moved or counting the dead rather than fabricating people entirely. More commonly, however, fabrication stems from structural misalignments between street-level implementation and the state data project’s goals. As studies of “bad data” in other organizations have shown \cite{Garfinkel_Bittner_1967, pine_institutional_2014}, internal organizational politics and mandates often shape the conditions that make fabrication both rational and necessary. Census enumerators have submitted falsified data due to piecework compensation structures that incentivized quantity over accuracy, or simply because fabrication required less effort than knocking on doors to survey residents. 

Context also matters. In one interview, an enumerator admitted to the second author that she inserted fake data when the software prevented her from recording exactly what a respondent said. She regularly encountered people who insisted that they are Puerto Rican\textemdash not Hispanic or white. She chose to mark those respondents as white and Hispanic simply because race and Hispanic origin are required data fields; she added Puerto Rican as a sub-category of Hispanic because this is permissible in the interface. She was not alone. When respondent information fails to align with predetermined categories\textemdash or when interface design leads to selection errors\textemdash government workers may produce data that do not fully reflect what they were given.

In Meritown, the first author saw a similar phenomenon, but the context and motivations shaped the dynamics differently. Most documented volunteering activities \textit{did} occur, but their recorded characteristics often deviated in degrees from what actually happened. Activities rarely lasted the mandated 90 minutes, and few met the minimum participation thresholds required by the municipal government. Rather than bending standards to fit reality, bureaucrats often found it more efficient to adjust reality on paper to fit data expectations. Bureaucrats compensated by inflating activity duration, overstating participant numbers, or recycling images from other events to make up data. The resulting data thus blended on-the-ground events with targeted alterations designed to align with municipal standards, producing hybrid data that served the quota-driven logic of higher-level supervision. 

Bureaucrats developed various techniques to manufacture or manipulate data. Consider the volunteer group photograph, a key piece of evidence used by the municipal government to validate activities. Because assembling 15 volunteers simultaneously was often infeasible, bureaucrats frequently included non-volunteers—passing pedestrians, committee staff, and sometimes the first author himself—in the group shots. When large groups happened to gather at the community committee offices, staff seized opportunities to stage multiple photographs to create a repository for future use. 

\begin{figure}[h]
    \centering
    \includegraphics[width=1\linewidth]{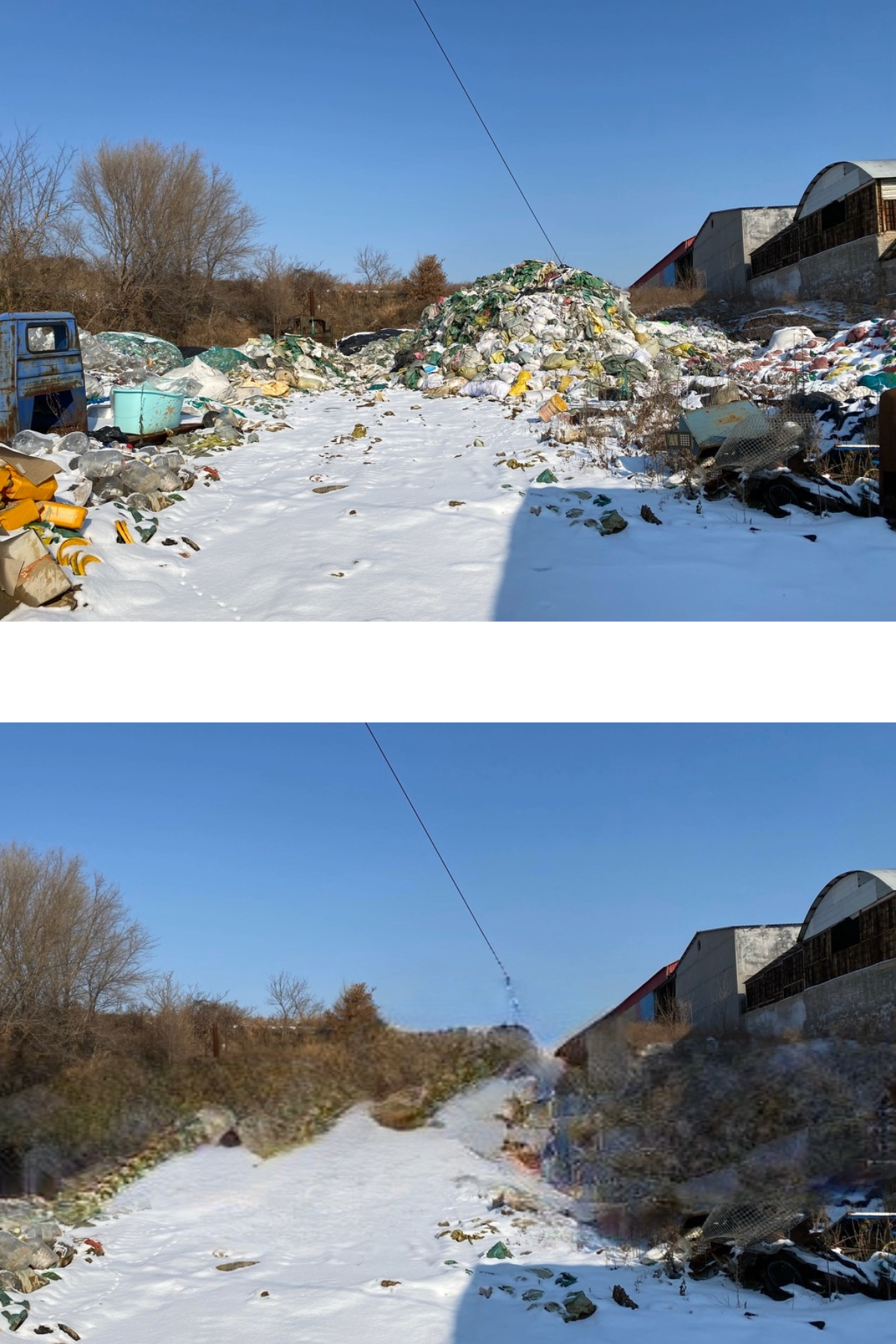}
    \caption{Comparison of images of the original village dump scene (Top) and the processed image after object removal (Bottom)}
    \label{fig:removel}
\end{figure}

Digital technologies have further enabled data fabrication. During the COVID-19 pandemic, Meritown required all volunteers to wear masks in photographs. Yet the absence of positive cases in the city led to widespread non-compliance. Instead of enforcing mask-wearing, bureaucrats routinely used photo-editing apps to digitally add masks. AI tools also became part of their repertoire. In November 2021, one grassroots committee discovered a mobile app capable of removing unwanted objects from images. Rather than organizing volunteers to clean neighborhood trash, staff simply deleted the garbage digitally (Figure \ref{fig:removel}), and uploaded the edited images as proof of a completed activity.

The motivations underlying the creation of fake data range from the explicitly political to the mundanely practical. Unintentional fakeness—often acknowledged by data creators—also emerges from misinterpreted instructions, data entry mistakes, or systemic limitations. Some of this fakeness is conceptualized as “error” but other aspects of created fakeness stem from the sociotechnical configuration itself.

\subsection{Correcting Fakeness}
Once created, data ascend through state bureaucratic hierarchies where they undergo various forms of processing, validation, and transformation. As data travel, their meaning and perceived fakeness shift. Each level of review involves actors who understand the data's limitations and their role differently. Correction is designed to better align data with reality according to the logics of the bureaucrats. But it is less a simple process of restoring accuracy than a process of realigning data with bureaucratic expectations.

The validation mechanisms differ between our two cases. In Meritown, township government officials validate volunteering reports primarily through visual evidence: counting heads in group photographs and verifying them against participant lists. By contrast, the census employs far more sophisticated procedures under the umbrella of “quality assurance,” checking for duplicate submissions, suspicious patterns, and logical inconsistencies. Notably, the census treats problematic data as routine and expected. Error rates are incorporated into the bureau's public reporting and other institutionalized practices. The system focuses on addressing problems with the data, regardless of their provenance; it only attends to systemic root causes when it can address them systematically (such as by creating new procedures for future enumerators). 

In Meritown, when township officials discover discrepancies in photographic evidence, they typically return cases to street-level bureaucrats, requesting a revision. Street-level bureaucrats may upload entirely new photographs with the correct number of participants, often with different people. Similarly, in the US census, enumerators may be dispatched to re-collect questionable data. However, bureaucrats sometimes edit data directly in lieu of demanding re-collection, acknowledging the cost and inevitability of honest mistakes. For example, census processors encountering a household with a 3-year-old “parent” of a 37-year-old presume a data entry error and reverse the relationship designation without seeking re-collection.

What constitutes data that require correction remains fluid and context-dependent. The census’s long-lasting handling of same-sex marriages illustrates this temporal contingency. For decades, when two household members identified as “male” and “married,” processors historically relabeled one as “female.” Following the legalization of same-sex marriage, the census modified this practice. Yet new complications emerged: an implausibly high number of same-sex married couples appeared in the data. Civil servants described how their investigation revealed that respondents often selected “male” and “married” by default, as those labels appeared first for each category in the interface. The census then developed algorithms to selectively edit responses, correcting some while preserving others, to produce statistically plausible distributions. Whether specific entries were “fake” versus “real” was less important than ensuring that the statistical portrait reflected the population.

Correction is also embedded in the power relations of the state's hierarchical structure, which can retroactively redefine whether a set of data is real or fake. When Meritown prepared for a provincial inspection in June 2021, inspectors demanded signature sheets that the city’s system had never required. This new standard suddenly rendered five years of otherwise “real” data “fake” in the eyes of  the provincial government. Rather than acknowledging this systemic oversight, bureaucrats spent weeks forging retroactive sign-in sheets, creating signatures to make their genuine activities legible to provincial standards. This episode shows that correction is not always about bringing data closer to reality because correction\textemdash and data processing practices more broadly\textemdash is never merely technical but always fundamentally political.

\subsection{Colluding Fakeness}
Although data correction may suggest otherwise, mid-level government workers are not always in an antagonistic relationship with street-level bureaucrats. Sometimes, they also collude with street-level bureaucrats in producing fake data to serve institutional needs while maintaining systemic coherence. 

This collusion often masquerades as correction within institutionalized data pipelines. Reconsider the AI-processed picture in the first section (Figure \ref{fig:removel}) . It appears plausible at first glance but quickly breaks down under closer scrutiny: the algorithm fills erased areas with distorted textures and unnatural edges. Then, consider when a street-level bureaucrat’s volunteer-activity data are rejected because a photo shows too few participants; she simply uploads another image with the “correct” number. In both cases, the supervisors who validate data are hardly naïve. But they will let the data pass. This is a coordinated collusion, producing data that meet political expectations rather than representational accuracy.

Such complicity emerges because fake data, contrary to common assumptions, are not failed data. Instead, they can be both necessary and productive \cite{lampland_false_2010}. Within Meritown’s political ecology, the objective is not accurate documentation of volunteer activities but rather the production of impressive statistics demonstrating the successful implementation of moral civilization campaigns. Township officials understand that inflated numbers that require fabrication serve this purpose more effectively than precise counts, provided they remain plausible enough to avoid triggering suspicion from even higher authorities. In this way, there is a “normalization of deviance” at work \cite{vaughan_dark_1999}.

A revealing incident during an unexpected inspection visit illuminates this dynamic. On a morning in September 2021, when a township official arrived at an urban grassroots committee office in Meritown to verify a volunteer activity that had been reported as currently in progress, she discovered that the volunteers had already departed. Rather than formally censuring this misrepresentation, the township official admonished the bureaucrat while clarifying her position:

\begin{quote}
“It is understandable that there are some mistakes in the event. I will not bother you if the event is not too abnormal. But remember, I am not the problem. If the supervisor sampled your problematic data, then you are done. They used to say they’ll inspect five committees per month; it is me that convinced them to only do two. Let’s just say, fake something that is more convincing!”
\end{quote}

This official explicitly endorses fabrication while articulating its parameters: data must be “convincing” enough to sustain the internal consistency logic for a quick skim by distant politicians and inspectors who assess success through decontextualized metrics rather than ground-level realities. Her role is not merely quality control; it is the active management of credible fiction: maintaining quotas, ensuring political priorities manifest in aggregate statistics, and smoothing suspicious patterns. Only in this way can statistically plausible narratives of policy success be maintained and promoted. The goal is not the representational accuracy of individual records, but a balance between the stable production of massive data and their plausibility. From this orientation, the link between data and reality becomes secondary; what matters is that the data appear internally consistent and aligned with institutional expectations.

Yet this systemic fabrication exacts psychological costs. The routine production of fake data erodes bureaucrats’ confidence in all state-produced information. As Chao, a township bureaucrat, once lamented:

\begin{quote} 
“They faked here, who knows what else they faked? […] I am so tired of faking things, but I have nowhere else to go. Faking data is the only skill I have now. Where can you find a place other than the government that needs this?” 
\end{quote}

This testimony shows how daily participation in data fabrication creates a self-perpetuating cycle. Bureaucrats internalize fake data production as a normal practice. For many reasons, they continue to create and work with fake data even when accurate documentation might be feasible. In the process, they construct a system where the production of fake data becomes both institutionally necessary and, potentially, personally corrosive.

Collusion is not always successful. Not all bureaucratic efforts to produce fake data to address political expectations are embraced. Collusion between some actors in a data-making configuration can create vulnerabilities at other levels. Consider the Census Bureau's multi-decade effort to implement a technique known as “statistical adjustment“ \cite{Belin_Rolph_1994} to address long-standing and consistent differential undercounts of specific sub-populations (including Black and Hispanic families, Native Americans living on tribal lands, and young children) \cite{O’Hare_2019}. The bureau's goal was not to increase representational accuracy at the individual data record level, but statistical accuracy so as to ensure that the allocation of political representatives and funding would be more equitable. To accomplish this, it sought to enroll a wide range of stakeholders into a shared project of producing data that were, at the individual record level, fabricated, yet statistically more truthful.

Initially, the statistical adjustment proposal triggered curiosity and confusion, as different stakeholders attempted to understand how this approach would impact their interests in the data. Yet, once Republicans began claiming that statistical adjustment was going to create fake data to benefit Democrats, broad support disintegrated \cite{sarathy_statistical_2024}. Through politicization, the collusion of different actors to leverage fakeness to improve data collapsed. 

Our comparison underscores that collusion around fakeness is only stable when institutions can maintain control over how the fakeness is framed and interpreted. In both cases, the legitimacy of adding fakeness into dataset hinges less on their representational value than on whether the boundary work can be successfully managed. And yet, the vulnerabilities opened through collusion reveal just how precarious this configuration can be. After all, fake data are only acceptable when their fakeness is not used to undermine their legitimacy.

\subsection{Augmenting Fakeness}
State actors also actively augment data in ways that further distance the data from a representational orientation. Some interventions are intentionally designed to undermine the system or achieve a different agenda. This aspect of fake data has been well studied in scholarship on statistical manipulation \cite{aragao_many_2022,wallace_seeking_2022}. However, other forms of augmentation arise less from malice than from the cumulative complexity of bureaucratic data work, where long chains of processing\textemdash involving multiple actors, media, and platforms\textemdash multiply opportunities for distortion.

In Meritown, for example, most of the street-level bureaucrats in villages are often in their 60s and lack the technical literacy required to complete digital datafication. They often document volunteering activities on handwritten paper sheets, which are handed to township bureaucrats to digitize. Errors routinely occur: misread handwriting, skipped lines, or mismatched names. Intentional or not, these practices push the resulting data even further away from the realities they are meant to represent.

Over time, state workers across levels also develop shared understandings about what forms of “fakeness” are acceptable, even necessary, for the system to function under resource constraints. They well understand these limits and often produce data with imperfections in mind, resulting in many data processing practices that may appear to make data “faker” on one level while “realer” at another.

The pursuit of “realer“ data through fakeness becomes especially clear in the US Census. The census’s mandate is to produce population statistics that support resource allocation and political representation while protecting confidentiality. With this goal in mind, the census accepts that individual records may not be an accurate reflection of the associated person so long as the data have overall statistical integrity. This dynamic is on display when the census must address missing data that enumerators were unable to collect. In such cases, the census uses a procedure known as “imputation” that leverages other data sources and bureaucratic processes to fill in the gaps. Political opponents of this method label the resultant data “fake” in order to undermine the legitimacy of the bureau’s work \cite{sarathy_statistical_2024, offenhuber_shapes_2024}.

After the 2000 census, the state of Utah sued the Census Bureau to block the use of imputation \cite{Cantwell_Hogan_Styles_2004}. In documents prepared for the Supreme Court, a senior demographer at the Census Bureau highlighted how the act of not imputing was statistically equivalent to imputing “zero” for a household, thereby knowingly creating an undercount. The Supreme Court ruled in the bureau’s favor \cite{breyer_utah_2002}, but opposition to this method persists \cite{boyd2026_data_made}.

Not all statistical endeavors are politically fraught. For example, those used to prevent “age heaping” are widely embraced. If the census were to publish age data as collected, there would be significantly more people with ages ending in 5 or 0 than other numbers; this is known as age heaping. Age heaping is both widespread and common every decade. Because the Census Bureau publishes individual census responses after 72 years, we can examine the discrepancy between reported ages and published statistics from 1880-1950. The original records show that age heaping occurred in each census. And yet, the published statistical data suggest otherwise. This is because the census “smooths“ the age data to prevent statistics that would appear peculiar.  Bureaucrats edit individual records, rendering them individually inaccurate while ensuring that the aggregate portrait reflects a reasonable distribution. Put another way, government workers edit “real“ data\textemdash effectively making them “fake“ data\textemdash so as to make the statistics “realer.”  

Across both the Meritown and census cases, we see the same epistemic trade-off: bureaucrats knowingly inject elements that make individual data points “fake” in order to achieve what they view as a more functionally real representation at the aggregate level. These practices reveal that fakeness is not simply a failure of representation but an institutionalized technique for managing uncertainty, aligning data with political expectations, and sustaining the operational legitimacy of state systems.

\section{Discussion}
As statistician George Box famously put it, “All models are wrong, but some are useful.” In a similar spirit, we propose that “all data are fake, but some are useful.” This provocation is not a nihilistic claim about truthlessness but an invitation to unsettle the binary thinking that positions “real” and “fake” data as opposing categories. Rather than imagining data as either accurate reflections of an external, stable world or corrupted deviations from it, our findings show that \textit{all} state data occupy a spectrum of divergence from the realities they purport to represent. More fundamentally, we challenge the assumption that data’s primary function is to mirror reality. Across both cases, “fakeness” emerges not as a defect in otherwise accurate records, but as a property produced and negotiated through sociotechnical practices. Fake data are a routine outcome of how bureaucrats, infrastructures, and institutional demands interact.

Crucially, the designation of data as “fake” depends on relational standpoints and social contexts. What appears as fake from one vantage point may be irrelevant, necessary, or unavoidable from another. Indeed, the very purpose of some data practices\textemdash such as statistical imputation or the correction of age-heaping\textemdash is to produce individually inaccurate records in order to ensure aggregate accuracy or population-level representativeness. This relational view is central to rethinking state legibility projects \cite{scott_seeing_1999}. It is also critical for moving away from the imagination of the state as “seeing” from everywhere and nowhere all at once. It is also at the heart of debates regarding efforts to produce synthetic data to ensure that statistical work can proceed without harming individuals \cite{whitney_real_2024,offenhuber_shapes_2024}. Understanding how states see demands attending to the historical, political, and organizational contexts in which state actors produce, interpret, and act upon data. These considerations ultimately raise central empirical and theoretical questions: fake for whom, in which situations, at what scales, and for what purposes?

Adopting a processual perspective further highlights how state data emerge from continuous contestations, compromises, and adjustments by diverse stakeholders across bureaucratic levels. Rather than focusing exclusively on the fake data’s moment of origin, we must analyze how data are linked, edited, combined, disaggregated, and reinterpreted as they travel through organizational pipelines. \cite{matzner_beyond_2016,feinberg_design_2017}. Bureaucrats are not naïve believers in perfect representation but sophisticated navigators of competing demands, balancing multiple pressures and objectives throughout the data lifecycle. Each stage presents opportunities for correcting certain representational inaccuracies while potentially introducing new ones. These transformations are political accomplishments, dependent on forms of expertise and legitimacy that “operates through the mechanisms that bring a debate to an end while keeping its potential continuation in sight” \cite{eyal_crisis_2019}. State fake data production, therefore, is best understood not as an unidirectional collection but as an ongoing negotiation. When examining “fake data,” specificity matters: at what stage of production it occurs, who processed it, through which methods, and according to what epistemic commitments?

From this view, the status of data as “fake” or “real” does not inhere in the record itself; it is enacted through practices of identification, evaluation, and mobilization. This is the performative perspective we adopt. Through our analysis, we show how labeling data as “fake” becomes a consequential moment that reveals shared or divergent understandings of institutional norms. Like infrastructure breakdowns that make hidden systems visible \cite{bowker_sorting_2000, singh_seeing_2021}, accusations of “fake data” disrupt routine processing and draw attention to otherwise opaque infrastructures of classification and authority. Such labeling practices extend beyond epistemic classification to encompass political action. Actors deploy the “fake data” label strategically to achieve specific goals, often connected to efforts at delegitimization. In some contexts, the designation matters little for meaningful social action, rendering the labeling practice itself meaningless. The act of observing, using, and labeling fake data is not separable from—but co-constitutive with\textemdash data’s “fakeness.” 

Taken together, our findings suggest that the key analytic question is not what fake data are but what the fakeness of data does. Indeed, “fake data” are sometimes more useful for achieving institutional ends than data that are ostensibly accurate. In China, fabricated volunteering records enable municipal governments to demonstrate compliance with political mandates even when underlying infrastructures fall short. In the US, imputed or edited census data produce more meaningful population-level estimates than publishing raw data would. Perceived usefulness thus shapes not only how agencies handle fake data ex post but also how they produce them ex ante: stakeholders’ anticipation of utility structures state data production from the outset. Yet when institutional expectations diverge from bureaucrats’ situated understandings of their work or of the data’s purpose, the result is exhaustion, cynicism, and diminishing trust within the state itself. The moral damage from routinized data fabrication is well documented in China \cite{zhao2023overstretched, chen_maintainers_2023} and is a source of political anxiety in the contemporary US \cite{Freilich_Kesselheim_2025}. Other countries are likely not to be immune as data-driven governance expands and institutional independence erodes.

Power shapes the dynamics of data fakeness in consequential ways. Not all actors hold equal authority in labeling data as “fake,” determining which processes are necessary, or deciding how data should be mobilized. In Meritown, for instance, street-level bureaucrats fabricate records to satisfy rigid metrics, yet township officials ultimately decide whether those fabrications are acceptable or even useful. Reconfiguring fakeness is, therefore, always a reconfiguration of power. Yet these dynamics are not purely top-down. They are often co-constituted by the public, who are also critical stakeholders in data production. To access benefits or evade surveillance, citizens routinely standardize, omit, or strategically manipulate biographical details to fit institutional categories \cite{liu_seeing_2022,singh_seeing_2021}, bringing their own interests and interpretations into the data-making process. 

The state has never required perfect information; it has always relied on “good enough” data to sustain administration. Yet the current moment introduces new pressures. Governments around the world increasingly promote techno-solutionist visions—from predictive policing to smart cities—that promise seamless representation and comprehensive monitoring \cite{vetro_data_2021,ammitzboll_flugge_street-level_2021,gordon_data_2024,winthereik_data_2024}. These sociotechnical imaginaries reshape public and political expectations, reviving the familiar trap of high-modernism in new high-tech form \cite{scott_seeing_1999,fourcade_learning_2020}. As state–society interactions become more thoroughly data-mediated, the discursive space for acknowledging the inevitable imperfections that enable governance shrinks.

Bureaucrats who understand the constraints and complexities of state data work now face mounting tension between technological hype and administrative reality. Unrealistic expectations generate pressure to fabricate for the sake of performing system functionality, echoing well-documented patterns of goal displacement in other large hierarchical organizations, such as hospitals \cite{pine_institutional_2014}, schools \cite{crooks_representationalism_2017}, newsrooms\cite{Christin_2022}, and tech platforms \cite{shestakofsky2024cleaning}, in which metrics become ends in themselves. At the same time, they may face intensified scrutiny for maintaining the very practices that keep the system operational but do not align with representational ideals. In this context, when citizens—or political actors like DOGE—encounter the state’s “fake data” with binary thinking, they may leap too quickly to accusations of fraud without understanding the institutional processes that produced these divergences. Such encounters exacerbate the ongoing crisis of expertise and deepen the erosion of trust in public institutions  \cite{eyal_crisis_2019,vertesi_resource_2023}. 

\section{Policy and Design Implications}
Our analysis challenges conventional approaches to data governance that frame “fake data” as a technical failure awaiting technological or policy solutions. The uncomfortable truth emerging from our work is that perfectly accurate state data remains both unattainable and potentially counterproductive within the structural constraints of the state. This is not a call to relax scrutiny. It is a call for contextual scrutiny that demands more critical attention to state-produced data. 

For responsible government agencies and citizens, the central question shifts from “How do we eliminate fake data?” to “How do we make responsible decisions with the imperfect information institutions can actually produce?” What matters is not maximizing abstract accuracy but governing necessary imperfections so they serve public purposes rather than narrow administrative interests, using frameworks that manage necessary uncertainties while maintaining democratic accountability.

\subsection{Toward Contextual Data Governance}
Similar to calls for contextual data governance in other fields \cite{liu_social_2022,winthereik_data_2024}, state data quality must be judged relative to intended use, collection constraints, and institutional context. State data governance should therefore adopt fit-for-purpose standards instead of universal accuracy thresholds. Evidence requirements should scale with decision risk. Eligibility adjudication or sanctions may demand tighter individual-level verification, whereas aggregate statistical reporting and policy planning may prioritize population-level representativeness and timeliness. 

Meanwhile, provenance and revision transparency are essential. Just as critical data scientists call for better data documentation in an era of AI \cite{Datasheets_2021}, public-sector data systems should document concepts, categories, and workflows; label records with provenance; and preserve auditable trails that retain dissent rather than bury it. Because political priorities and social classifications evolve, governance frameworks should include periodic reviews of quality standards, with public rationales for changes and guidance on their implications for interpretation. At the same time, we acknowledge that\textemdash and need to more actively contend with\textemdash the ways in which transparency is often itself politicized in the government context \cite{Pozen_2018}. Likewise, we recognize that state actors struggle to communicate uncertainty and statistical limits to public audiences \cite{Kreps_Kriner_2020}. More work is needed to bridge the epistemic gaps that surround government data and statistics \cite{boyd_differential_2022}. 

Lastly, agencies should make explicit the political and administrative logics behind editing protocols and quality thresholds. Treating data quality as purely technical obscures choices that are, in fact, normative. Data never “speak for themselves” \cite{bowker2008memory}. Agencies should pair releases with interpretive memos that explain purposes, trade-offs, uncertainties, and known limitations. Adding narratives to data is not an extraneous add-on that threatens legitimacy; it is integral to governance. At the same time, explanations about data-making can easily be weaponized, as they have been in domains like climate science, vaccine policy, and public health \cite{Proctor_Schiebinger_2008}.  This is perhaps the greatest risk in data communication—and much more work in this area is necessary.

\subsection{Uncertainty-Forward System Design}
Echoing other HCI researchers, we argue that state data interfaces should foreground uncertainty and provenance as essential features, making visible the institutional contexts and transformation processes that shaped particular data. This call builds on seamful design, which foregrounds the limitations and boundaries of artifacts rather than smoothing them away \cite{Chalmers_MacColl_Bell_2003,Inman_Ribes_2019}, as well as adversarial design, which highlights the conflicting politics embedded in sociotechnical systems instead of concealing them \cite{disalvo2015adversarial}. In a similar spirit to Seberger and Gupta’s critique of the “data double” \cite{seberger_designing_2025}, fake data should not be understood as a distorted replica of reality striving to close the gap towards sameness, but as its own entanglement—one that foregrounds divergence, situatedness, and difference rather than aspiring to perfect correspondence.

Designing for uncertainty continues to be an open question in HCI \cite{Greis_Hullman_Correll_Kay_Shaer_2017,Soden_Devendorf_Wong_Chilton_Light_Akama_2020,Giaccardi_Murray-Rust_Redström_Caramiaux_2024}. However, we offer a few design commitments in light of our analysis. Rather than reinforcing binaries, interfaces should help data users identify the messiness of data, perhaps by displaying confidence measures alongside data points and specify provenance or distinguishing between direct observations, statistical imputations, and policy-driven synthetic entries. Analyses of data should capture and preserve dissent rather than surface only consensus, perhaps by maintaining records of alternative interpretations that might prove relevant for audit or review. Another approach would be for interfaces to support graduated\textemdash rather than binary\textemdash quality indicators, enabling users to understand the spectrum of confidence associated with different data points. That said, much more research is needed to determine how to present uncertainty without unintentionally misleading data users \cite{Hofman_Goldstein_Hullman_2020}.

Creating and sensemaking within such uncertainty-forward systems imposes a burden for both data users \cite{Hullman_2020} and overworked bureaucrats to manage within resource constraints. Yet, the alternative is not a system free of this burden, but one where the labor of managing uncertainty remains hidden and illicit, reinforcing the notion that there is “real” data and “fake” data. For HCI practitioners, the goal should be to formalize the inevitable labor of uncertainty so that it shapes democratic accountability. By capturing dissent and preserving edit histories, systems can render the politics of fictions legible, enabling decisions that are transparent, contestable, and commensurate with public purposes rather than narrow administrative interests.

Democratic oversight should focus less on purging all imperfections and more on ensuring that the imperfections we tolerate advance public aims and distribute risks fairly. However, it is also imperative to contend with how statistical uncertainty is often pejoratively interpreted by political actors as fake \cite{Steed_Liu_Wu_Acquisti_2022} and, thus, can reinforce the very problems that designing with uncertainty seeks to remedy. Moreover, uncertainty has been\textemdash and will continue to be\textemdash leveraged by those seeking to delegitimize state data \cite{Edwards_2013,Proctor_Schiebinger_2008}. This is all to say that context matters.

\section{Conclusion}
This paper traced the production of data fakeness through bureaucratic pipelines in two distinct contexts: Chinese volunteering documentation and data processing in the 2020 US Census. By examining how fakeness is created, corrected, colluded, and augmented across bureaucratic hierarchies, our ethnographic analysis reveals that what gets labeled as “fake data” is not a fixed category of failed representation but rather an ongoing accomplishment negotiated among multiple actors with divergent objectives, constraints, and institutional contexts. We develop a relational, processual, and performative framework for understanding state data that enables critical engagement with state data practices while avoiding both naive realism and skeptical nihilism. We close by returning to our provocative claim: all data are fake, but some are useful. This is not cynical resignation but pragmatic recognition. The lesson is to be specific and even boring\cite{star1999ethnography}; to work with state data, we must be willing to accept the messiness and compromised reality of how these data are produced. Doing so requires skepticism toward high-modernist promises and data solutionism that can never be fully realized.

\begin{acks} 
We would like to thank our anonymous reviewers for their fantastically helpful suggestions. We had the opportunity to workshop earlier drafts of this work at three academic meetings\textemdash Social Science History Association, Society for Social Studies of Science (4S), and American Sociology Association\textemdash where attendees helped us sharpen our thinking. Liu's work was supported by an American Sociological Association Doctoral Dissertation Research Improvement Grant and the Horowitz Foundation for Social Policy. The Alfred P. Sloan Foundation (G-2019-12414), the John S. and James L. Knight Foundation, and the National Science Foundation (2429838) provided support for boyd's ethnographic work. We are especially grateful to Kevin Ackermann, Parker Bach, Nancy Baym, Dan Bouk, Yuchen Chen, Tarleton Gillespie, Mary Gray, Martha Lampland, Jayshree Sarathy, Ranjit Singh, Ryland Shaw, Irina Shklovski, Emily Tseng, and Janet Vertesi for helping us think through the ideas in this paper. 

Finally, we want to acknowledge the invisible government bureaucrats who opened up to us and graciously allowed each of us to learn from them. We hope that our paper helps others understand what they do and why they do it.
\end{acks}

\bibliographystyle{ACM-Reference-Format}
\bibliography{fake}

\end{document}